# Coherent control of flexural vibrations in dual-nanoweb fibers using phase-modulated two-frequency light

J. R. Koehler[1,*], R. E. Noskov[1], A. A. Sukhorukov[2,1], D. Novoa[1], and P. St.J. Russell[1]

[1]*Max-Planck Institute for the Science of Light, Staudtstr. 2, 91058 Erlangen, Germany*
[2]*Nonlinear Physics Centre, Research School of Physics and Engineering,
Australian National University, Canberra, ACT 2601, Australia*

**johannes.koehler@mpl.mpg.de*

Coherent control of the resonant response in spatially extended optomechanical structures is complicated by the fact that the optical drive is affected by the backaction from the generated phonons. Here we report an approach to coherent control based on stimulated Raman-like scattering, in which the optical pressure can remain unaffected by the induced vibrations even in the regime of strong optomechanical interactions. We demonstrate experimentally coherent control of flexural vibrations simultaneously along the whole length of a dual-nanoweb fiber, by imprinting steps in the relative phase between the components of a two-frequency pump signal, the beat frequency being chosen to match a flexural resonance. Furthermore, sequential switching of the relative phase at time intervals shorter than the lifetime of the vibrations reduces their amplitude to a constant value that is fully adjustable by tuning the phase modulation depth and switching rate. The results may trigger new developments in silicon photonics, since such coherent control uniquely decouples the amplitude of optomechanical oscillations from power-dependent thermal effects and nonlinear optical loss.

## I. INTRODUCTION

The technique of "coherent control" is attracting increasing interest because it uniquely allows one to manipulate the coherent response of mechanical or optical degrees of freedom in a wide range of physical systems. So far, experiments have relied on electrical driving of coupled micromechanical resonators [1–3] and also on driving of nanophotonic objects using optical fields with spatio-temporally shaped amplitude and/or phase [4]. For example, trains of phase-locked ultrashort pulses, spaced by a period smaller than the dephasing time of the excitations and often assisted by spectral or spatial phase-shaping, have been used to trigger photochemical reactions along selected pathways [5–9], to control light localization in plasmonic nanostructures [10], and to observe vibrational wave-packets interfering in molecular ensembles [11] and even single molecules [12]. Coherent control using pulse trains has also been applied to inelastic light scattering, finding applications in coherent anti-Stokes Raman spectroscopy [13–16], selective excitation of optical phonons in molecular crystals [17,18] and optoacoustical effects in the micron-scale core of photonic crystal fibers (PCFs) [19].

PCFs share an important feature with structures as disparate as silicon nanowires [20,21] and dual-nanoweb fibers [22]: they are all key examples for a class of microstructured optical waveguides characterized by a giant acoustic impedance mismatch around the light-guiding region [23]. For this reason, they support, apart from acoustic modes with a purely longitudinal wavevector, also entirely transverse vibrations closely resembling the internal degrees of freedom causing stimulated Raman scattering (SRS) in molecular materials: their characteristic frequency-wavevector relation (Fig. 1(a), bottom) is, similar to optical phonons in a diatomic lattice, flat close to the cut-off frequency $\Omega_R$ (i.e., $\partial\Omega/\partial\beta_{ac}\big|_{\Omega\to\Omega_R} \simeq 0$). In this region, the wavevector $\beta_{ac}$ can be freely chosen without changing the frequency, so that associated acoustic phonons can mediate transitions between optical side-bands that satisfy the frequency-wavevector relation of e.g. the fundamental optical mode (Fig. 1(a), top). When two mutually coherent laser signals are launched at the pump and 1st-order Stokes frequencies ($\omega_0$ and $\omega_{–1}$, marked by black filled circles), spacing equal to $\Omega_R$, their beat-note drives the generation of phonons through electrostriction or radiation pressure. Since $\Omega_R$ is much smaller than the optical frequency, optical group-velocity dispersion (GVD) is negligible and the same phonon can mediate transitions between many pairs of adjacent side-bands, giving rise to stimulated Raman-like scattering (SRLS) with a pump-frequency independent frequency shift $\Omega_R$. In contrast to SRS in free space, where phase-matching requires the side-bands to be non-collinear with the pump beam, limiting the interaction length, in SRLS the whole waveguide is "active" because all the side-bands are automatically and exactly phase-matched. If the amplitude or relative phase of the launched driving fields is changed, and the time-of-flight of an optical signal is much less than the acoustic period $(2\pi/\Omega_R)$, i.e.:

$$\Omega_R L\, n_G / (2\pi c) \ll 1 \qquad (1)$$

(*c* is the speed of light in vacuum, $n_G$ the group index of the light and *L* the waveguide length) then the resulting optical pressure will change instantaneously along the whole waveguide. For our experimental parameters ($L = 0.22$ m, $n_G = 1.4$, $\Omega_R/2\pi = 5.6$ MHz) the expression on the left is 0.006.

However, even if this condition is fulfilled, coherent control of acoustic vibrations in spatially extended systems with strong optomechanical coupling is often challenging. Indeed, it is commonly accepted that since the optical driving field experiences back-action from generated phonons, its associated pressure varies with position along the waveguide.

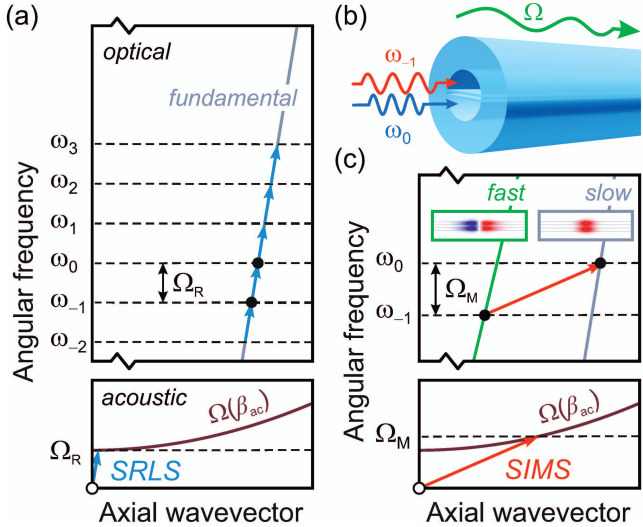

FIG. 1. (a) Stimulated Raman-like scattering (SRLS) by transverse acoustic vibrations: the upper and lower diagrams depict optical and acoustic dispersion schemes, an open circle indicating zero wavevector. Seeding an optical mode (e.g. fundamental, grey line) with two laser frequencies $\omega_0$ and $\omega_{-1}$ drives acoustic vibrations at their beat-frequency $\Omega = \omega_0 - \omega_{-1}$. Generated for $\Omega = \Omega_R$, identical SRLS phonons (blue arrows) mediate a cascade of intra-modal transitions. (b) Sketch of an idealized, axially uniform dual-nanoweb fiber: two thin glass membranes, spaced by a narrow gap, are mounted inside a capillary and optically driven to perform flexural vibrations. (c) Dispersion schemes for stimulated inter-modal scattering (SIMS) between two optical eigenmodes of the dual-nanoweb fiber: to generate SIMS phonons (red arrows) for $\Omega = \Omega_M$, $\omega_0$ is launched into the slow and $\omega_{-1}$ into the fast mode (transverse field distributions shown as insets). In contrast to SRLS, only a single inter-modal transition is possible.

Here we reveal a general condition under which this paradigm breaks down: the optical pressure related to SRLS always remains uninfluenced by the induced vibrations, even though strong optomechanical interactions may generate multiple optical side-bands. We realize this regime experimentally and demonstrate coherent control of optically-driven flexural vibrations in a dual-nanoweb fiber simultaneously over its whole length by modulating the relative phase of the two-frequency laser drive on a timescale less than the SRLS lifetime (typically ~160 μs). On the one hand, this permits continuous tuning of the vibrational amplitude without changing the optical power, in contrast to techniques varying the repetition rate of driving pulses [19]. On the other hand, we demonstrate using this technique, that the unique features of SRLS enable suppression of flexural vibrations simultaneously along the whole fiber length, while being robust against axial non-uniformities.

The employed dual-nanoweb fiber, sketched in Fig. 1(b), contains a nanostructured waveguiding region consisting of two closely-spaced and optically-coupled silica membranes, attached to the inner walls of a fiber capillary [24]. Optical gradient pressure deflects these mechanically highly compliant "nanowebs", giving rise to a large change in modal refractive index with launched power. By driving at a flexural resonant frequency, the optomechanical nonlinearity can be further enhanced, making it possible to detect the vibrations by transverse optical probing through the fiber cladding. This technique has revealed a type of optoacoustic scattering qualitatively different from SRLS to occur in the dual-nanoweb fiber: stimulated inter-modal scattering (SIMS) between the fundamental ("slow") and a double-lobed ("fast") optical mode [25], recently also investigated in hydrogen-filled hollow-core PCFs [26] and in silicon nanowires [27]. The dispersion scheme of this process is sketched in Fig. 1(c). In strong contrast to SRLS, only a single phase-matched SIMS transition is possible because the axial wavevectors of pump and Stokes photons in the two optical modes (Fig. 1(c), top) are wavelength-dependent and differ significantly. For this reason, SIMS transitions require a considerably larger wavevector than SRLS. Owing to the dispersion of the guided flexural vibration (Fig. 1(c), bottom), this results in a SIMS frequency shift $\Omega_M$ that is typically larger than $\Omega_R$. However, measurements of the spatio-spectral distribution of flexural vibrations along the fiber have revealed that at certain positions intra-modal SRLS and inter-modal SIMS coexist at the same frequency ($\Omega_M = \Omega_R$). Only possible if the resonant frequencies of the two nanowebs are different, this highly unusual situation proves that the dual-nanoweb fiber suffers from significant structural non-uniformities along its length, permitting the stricter condition for SIMS phase-matching to be occasionally met at certain locations. The resulting interplay of SRLS and SIMS allows the nanowebs to self-oscillate when pumped with only a few mW of monochromatic light [22]. This effect would not be possible in the absence of SIMS, because coherent suppression of pure SRLS gain is expected.

In the following section, a theoretical model is derived that demonstrates the unique advantages of coherent control through SRLS over SIMS. We employ this model to carry out numerical simulations of the spatio-temporal dynamics of both processes and also investigate the impact of axial fiber non-uniformities on coherent control.

## II. THEORETICAL MODEL AND NUMERICAL SIMULATIONS

Flexural vibrations of the nanowebs at frequency $\Omega$ are driven when two mutually coherent single-frequency laser signals, spaced by this frequency, beat against each other (see Fig. 1(b)). The complex amplitude of their beat-note, acting as an effective optical pressure, is given at the fiber input by [25]

$$\Phi_0 = s_0(z=0) s_{-1}^*(z=0), \quad (2)$$

where $s_n$ is the slowly-varying electric field amplitude of the $n^{th}$ side-band in TE polarization, with pump referred to as $s_0$ and Stokes as $s_{-1}$. Normalization and boundary conditions require $s_n(z=0) = [P_n(z=0)/P_{IN}]^{1/2}$, where $P_n(z=0)$ is the power launched into the $n^{th}$ side-band and $P_{IN}$ the total launched power.

If the beat-frequency of the two-frequency light matches to the SRLS frequency $\Omega_R$ in one nanoweb, power can be transferred between many equally-spaced SRLS side-bands as the light progresses along the fiber. Accounting for these, the SRLS pressure is in its general form expressed as

$$\Phi(z,t) = \sum_n s_n(z,t) s_{n-1}^*(z,t). \quad (3)$$

Since the geometrical parameters of the nanowebs (namely their widths, thicknesses, convex profiles and the interweb spacing) vary much more slowly along the axial fiber length than in the transverse direction, the SRLS pressure is related to the normalized slowly-varying complex envelope $R(z,t)$ of flexural vibrations by [25]

$$\left[\frac{\partial}{\partial t} + \frac{1}{L(z,\Omega)} + v_g \frac{\partial}{\partial z}\right] R(z,t) = i\kappa_R \Phi(z,t), \quad (4)$$

where $L(z, \Omega) = (2i\Omega)/(\Omega^2 - \Omega_R^2(z) + i\Omega\Gamma_R)$ accounts for the Lorentzian shape of the mechanical resonance, $\Gamma_R = \tau^{-1}$ is its linewidth and $v_g$ the group velocity of SRLS vibrations. The beat-note pressure couples to the vibrations at rate $\kappa_R = [\Gamma_R \Omega_R c\, g_0 / (2\omega_0)]^{1/2}$, where $g_0 \approx 0.2\ \mu m^{-1} W^{-1}$ is the peak value of the optomechanical gain factor for SRLS. In general, non-uniformities of the dual-nanoweb structure along the fiber will cause $\Omega_R$ to become position-dependent. For simplicity, this effect is disregarded here and will be treated separately in Section II C.

At each point in time, the envelope of the flexural vibrations and the slowly-varying amplitudes of the side-bands are coupled by the quasi-steady-state relation (since an oscillation cycle of optical fields is much shorter than a period of the flexural vibrations) [25]

$$\frac{\partial}{\partial z} s_n(z,t) + \frac{\alpha}{2} s_n(z,t) = i\frac{\omega_n}{\omega_0} \kappa_s \left[ R(z,t) s_{n-1}(z,t) + R^*(z,t) s_{n+1}(z,t) \right], \quad (5)$$

where $\kappa_s = \kappa_R \omega_0/(\Omega_R c_0)$ is the coupling rate between the optical side-bands per unit length of the nanowebs [25] and $\alpha$ the power loss of the fundamental optical mode.

It is most remarkable that, although flexural vibrations strongly reshape the optical spectrum through the generation of side-bands according to Eq.(5), the optically induced pressure is independent of the phonon amplitude and evolves along the fiber simply as [25]

$$\frac{\partial}{\partial z} \Phi(z,t) = -\alpha \Phi(z,t). \quad (6)$$

The underlying reason is that the phonons only modulate the phase of the optical components, while the optical intensity profile remains unchanged due to the absence of GVD over the side-band frequency range. Purely defined through the optical intensity and insensitive to phase, the SRLS pressure is only determined by the input optical spectrum and optical loss. Importantly, this result also holds in presence of axial non-uniformities along the fiber (see Section II C). It is this property of SRLS that enables simultaneous coherent control over flexural vibrations along the whole fiber.

### A. Analytic theory for coherent control of flexural vibrations through SRLS

We now consider coherent control of SRLS phonons by changing the relative optical phase of pump and Stokes with time by $\varphi(t)$ at the fiber input while keeping their powers constant. Under these circumstances, Eqs.(2) and (6) yield for the SRLS pressure $\Phi(z,t) = \Phi_0(z)\, e^{i\varphi(t)}$ (where $\Phi_0(z) = \Phi_0\, e^{-\alpha z}$). The induced spatio-temporal response of the flexural vibrations can be calculated by directly solving Eq.(4). Assuming the dual-nanoweb structure to be perfectly uniform along the fiber, we obtain for zero detuning between drive and SRLS frequency ($\Omega = \Omega_R$) and for zero group velocity of SRLS phonons ($v_g = 0$)

$$R(z,t) = i\kappa_R\, e^{-t/(2\tau)} \int_{-\infty}^{t} d\vartheta\, \Phi(z,\vartheta)\, e^{\vartheta/(2\tau)}. \quad (7)$$

For an abrupt phase step $\varphi(t) = \Delta\varphi\, \Theta(t)$ at time $t = 0$ in the two-frequency beat-note, where $\Theta$ is the Heaviside function, Eq.(7) simplifies as

$$R(z,t) = |R(z,t)| e^{i\Psi(t)}$$
$$= 2i\kappa_R \Phi_0(z) \tau \left[ (1 - e^{i\Delta\varphi}) e^{-t/(2\tau)} + e^{i\Delta\varphi} \right], \quad (8)$$

where ψ(t) is the phase-lag of the flexural vibrations relative to the drive. Hence, at a fixed position along the fiber the normalized phonon population $|R(z,t)|^2$, which equals unity at $t = 0$ and for $t \gg \tau$, evolves as

$$|R(t)|^2 = 1 + 4\left(e^{-t/\tau} - e^{-t/(2\tau)}\right)\sin^2(\Delta\varphi/2). \qquad (9)$$

For all values of $\Delta\varphi$, the phonon population reaches a minimum at time $t_{min} = 2\tau \ln 2$, at which point the phase-lag $\psi = (\Delta\varphi - \pi)/2$ has undergone half of its total shift to catch up with the drive. Although $t_{min}$ in Eq.(9) is a fixed quantity because it is solely defined by the SRLS lifetime, it can be reduced significantly if a sudden increase of the two-frequency power is synchronized with the phase-step. Theoretical details and their confirmation in a series of measurements are given in the Appendix.

We note that, although enabling a simple analytic solution, the Heaviside step function $\Theta(t)$ violates the slowly-varying envelope approximation used in the derivation of Eqs.(4) and (5). Therefore, Eq.(8) is formally invalid within a short interval around the time of phase-switching (during which the spectrum of the driving beat-note pressure significantly broadens). However, outside this time interval the system's response obeys to Eq.(8).

## B. Simulations of full spatio-temporal optomechanical nonlinear dynamics

In this section, we perform direct numerical simulations of the spatio-temporal optomechanical dynamics and compare the suppression of SRLS vibrations discussed above with the regime of SIMS, a close relative of conventional stimulated Brillouin scattering.

First, we present the results of our SRLS model. Accounting for the finite response time of the pump-Stokes phase-flip in the experiments, we approximate the step function with $\Theta(t) \approx \pi^{-1}[\arctan(t/\varepsilon) + \pi/2]$ and employ a typical experimental value $\varepsilon \approx 34$ ns. The full spatio-temporal evolution of optical side-bands and flexural vibrations can then be simulated with Eqs.(4) and (5), in which we for simplicity disregard optical losses and axial non-uniformities.

Figure 2 plots the optomechanical nonlinear dynamics in a 22 cm-long dual-nanoweb fiber: two-frequency light, with its beat-note tuned to $\Omega_R$, is switched on at $t = -9\tau$. This initiates a transfer of optical power from pump to Stokes that gradually grows both with time and position along the fiber (Figs. 2(a) and (b), accompanied by the generation of coherent SRLS phonons. A fraction of them is, in turn, immediately annihilated, causing a frequency-upshift of pump photons and the emergence of a relatively weak signal at the first anti-Stokes frequency (Fig. 2(c)), which follows the build-up of the SRLS phonons over time (Fig. 2(d)). Their population is independent of position along the fiber since in the absence of optical losses Φ is constant everywhere.

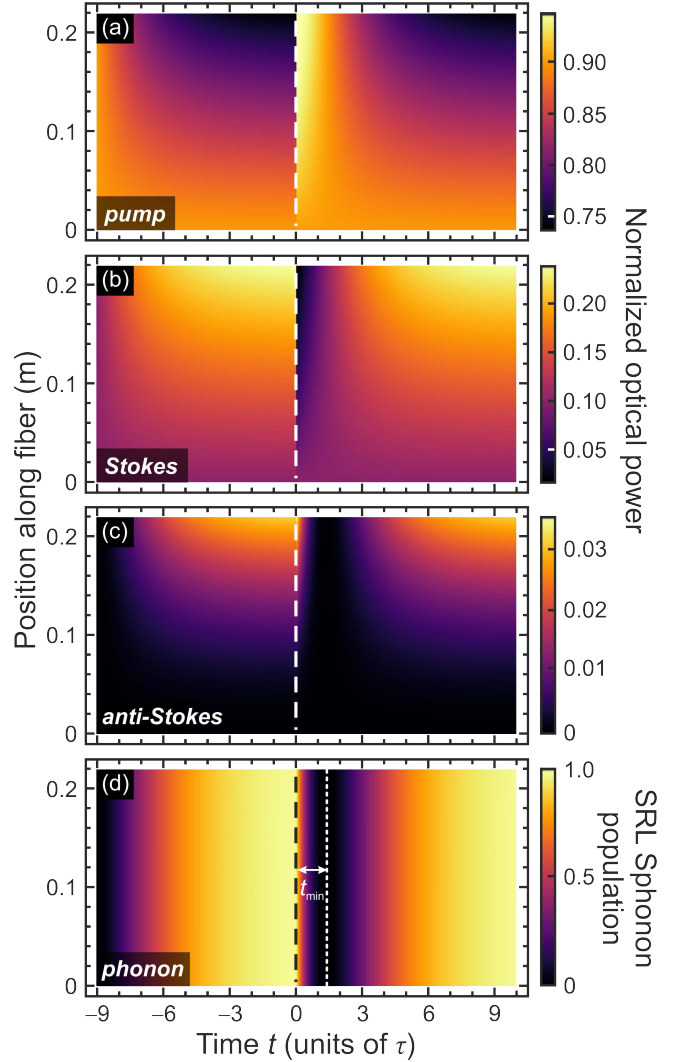

FIG. 2. Optomechanical nonlinear dynamics of SRLS, numerically simulated for a lossless uniform 22 cm-long dual-nanoweb fiber. Spatio-temporal evolution of the normalized optical power in the (a) pump and (b) Stokes side-bands: both signals are switched on at $t = -9\tau$, initiating a pump-to-Stokes power transfer. The emergence of the 1st-order anti-Stokes side-band (c) reflects the presence of a coherent population of SRLS phonons. Its spatio-temporal dynamics is shown in (d) where the steady-state value (reached before $t = 0$) is normalized to unity. The relative pump-Stokes phase is inverted at $t = 0$ (dashed line), causing a transiently reversed power flow from Stokes/anti-Stokes to pump while phonons interfere destructively. Their population vanishes simultaneously everywhere after time $2\tau\ln 2$ (dotted line). The simulations were carried out with pump and Stokes seed powers of 27 and 3 μW, respectively [28].

The optical powers and the SRLS phonon population reach their steady state after ~7 SRLS lifetimes, before at $t = 0$ the relative phase of pump and Stokes seed is flipped to invert the SRLS pressure. While the phonons generated in opposite phase before and after $t = 0$ interfere destructively, a transiently reversed flow of optical power from Stokes and anti-Stokes back to pump is observed. Agreeing with Eq.(9), the SRLS phonon population and its related anti-Stokes signal vanish after time $t_{min} = 2\tau \ln 2$ everywhere along the fiber (dotted line in Fig. 2(c)) and recover after that to their steady state.

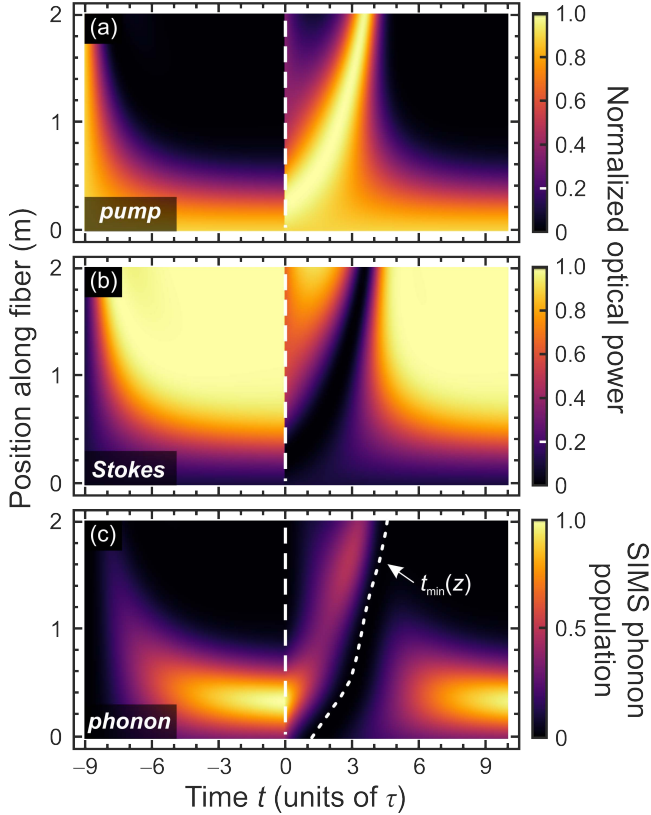

FIG. 3. Optomechanical nonlinear dynamics simulated along a lossless uniform 2 m-long dual-nanoweb fiber when SIMS drives flexural vibrations with a group velocity of 12 m/s. Spatio-temporal evolution of the (a) pump and (b) Stokes side-band powers: the two-frequency light is switched on at time $t = -9\tau$. Pump-to-Stokes power transfer leads to a simultaneous build-up of a SIMS phonon population, whose normalized evolution is plotted in (c). Annihilation of SIMS phonons begins after $t = 0$ (dashed line) when the relative pump-Stokes phase is flipped, accompanied by a transient Stokes-to-pump back-flow of optical power. In contrast to SRLS in Fig. 2, the time interval $t_{min}$ (dotted line) strongly depends on the position along the fiber. For these simulations, we employed pump and Stokes seed powers of 0.5 mW and 50 μW, respectively. The peak gain coefficient of SIMS was $g_0 \approx 0.05$ μm$^{-1}$W$^{-1}$ [25].

A fundamentally different situation emerges if the beat-note is tuned to the SIMS frequency $\Omega_M$ ($\sim 2\pi \times 150$ kHz larger than $\Omega_R$): associated phonons, propagating at a group velocity $v_g \approx 12$ m/s, can only phase-match a single transition between optical modes that are single- and double-lobed along the nanowebs. To visualize coherent control of flexural phonons in this regime, we consider a 2 m-long homogeneous dual-nanoweb fiber and disregard again optical losses. Figure 3 plots the numerically simulated spatio-temporal evolution of SIMS when (in analogy to Fig. 2) the two-frequency light is switched on at time $t = -9\tau$. In contrast to the SRLS case, the Stokes signal is amplified at the cost of a fully depleted pump (see Figs. 3(a),(b)) for $t < 0$ within the first 60 cm, co-localized with the major part of the SIMS vibrations (Fig. 3(c)). The transiently reversed transfer of optical power from Stokes to pump is directly after $t = 0$ also most pronounced in that region; the position where it is strongest subsequently experiences a temporal delay that increases with position.

However, the key observation is that in SIMS it is impossible to simultaneously suppress the flexural vibrations along the whole fiber. Figure 3(c) shows indeed that the phonon suppression time $t_{min}(z)$ gradually increases with position along the fiber: its trajectory, indicated by a dotted line, is curved, in strong contrast to SRLS. This is because in SIMS the flexural vibrations in a section of the fiber strongly influence the optical pressure acting on the subsequent section, i.e. Eq.(6) does not apply. Therefore, only after the phonons are suppressed in the initial fiber section, they gradually become suppressed further along the fiber. Additional simulations (not shown) confirm that this delay is enhanced by the group velocity of SIMS phonons, causing the position of maximum phonon population to shift away from the input fiber end.

### C. Impact of axial fiber non-uniformities on the optomechanical nonlinear dynamics

Next, we apply Eqs.(4) and (5) to simulate the impact of axial non-uniformities on the coherent control, taking into account the actual profiles of $\Omega_{R(M)}(z)$ measured in Ref. [25], which reveal maximum phonon-frequency variations of 1% (typically within ~6 kHz for SRLS), so that two-frequency light can drive nanoweb vibrations in those fiber sections where the beat-frequency matches the local SRLS/SIMS frequency, as sketched in Fig. 4(a).

Figure 4(b) plots the calculated spectral mean frequencies (relative to the optical pump frequency) of the SRLS and SIMS side-bands versus position, shading indicating the rate of frequency shift which is, in turn, a measure for the mechanical work performed by the optical fields. Pronounced red-shifts in the local optical spectrum, corresponding to shaded areas, coincide with the frequency-matching points in Fig. 4(a), indicating most efficient phonon generation at these locations. Perfect phase-matching between many adjacent optical side-bands allows the SRLS spectrum to acquire a total shift of $\Delta\Omega \sim 5\,\Omega_R$, whereas for SIMS the upper limit is $\Omega_R$ since only a single transition is possible. The spatial evolution of the SIMS beat-note pressure in Fig. 4(c) also reflects this; its maximum occurs at ~32 mm, where the powers in the pump and Stokes SIMS side-bands are identical.

However, the key result of these simulations is that, although structural non-uniformities cause the nanoweb deflection and the optical spectrum to vary with position, the SRLS pressure is, unlike for SIMS, constant along the whole fiber, as predicted by Eq.(6).

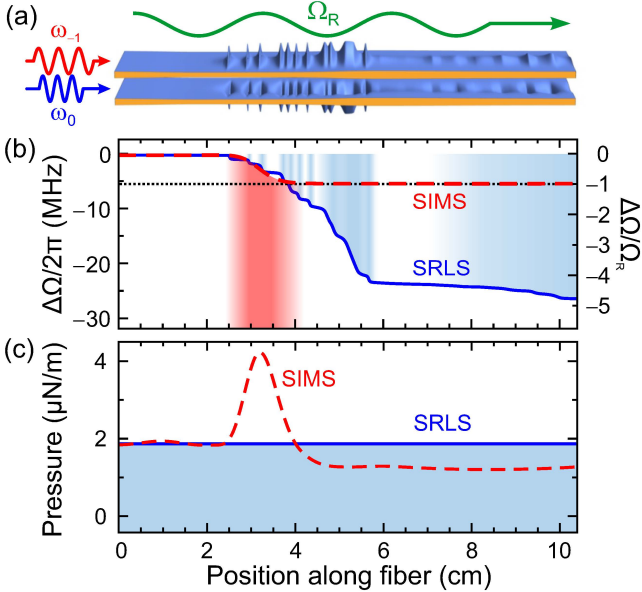

FIG. 4. (a) Typical spatial distribution of the anti-phase nanoweb peak deflection, simulated for SRLS vibrations driven with two-frequency light in an axially non-uniform dual-nanoweb fiber. (b) Calculated mean frequency of transmitted light; mechanical work done by the optical fields causes a net red-shift, shading indicating its rate. (c) Numerically simulated optical pressure (force per unit fiber length) for SRLS and SIMS, obtained by integrating over the nanoweb width. While the SIMS pressure (red dashed line) peaks at ~32 mm, the SRLS pressure (blue solid line) is after Eq.(6) constant along the whole fiber.

## III. EXPERIMENTAL SETUP

In view of the fundamental advantages outlined above, we focus our experiments on coherent control through SRLS. The scanning electron micrograph (SEM) in Fig. 5(a) shows the cross-section of the dual-nanoweb fiber, zoomed in around its waveguide region: the ~22 μm-wide nanowebs are ~440 nm thick at their center, and spaced by ~550 nm. Both ends of a 22 cm-long fiber sample (Fig. 5(b)) were mounted in gas cells and evacuated to ~1 μbar to suppress gas damping. This resulted in a mechanical linewidth of a few kHz [29]. A two-frequency signal was synthesized by down-shifting a portion of the output of a single-frequency laser at 1550 nm (~16 kHz linewidth) using a single-sideband modulator, passing the remaining unshifted light through an electro-optic phase modulator (PM) and combining them with a pump-to-Stokes power ratio of 9:1. The two-frequency signal was then amplified and launched (TE-polarized) into the nanowebs. The nanoweb vibrations will cause the appearance of equally-spaced higher-order SRLS Stokes and anti-Stokes sidebands in the transmitted signal. After heterodyning with a local oscillator (LO), these are detected using a photodiode (PD) and an RF spectrum analyzer (RF-SA) [22]. The temporal evolution of the SRLS side-bands can thus be tracked by monitoring the RF power at a selected beat-frequency with the RF-SA (~7.5 μs temporal resolution).

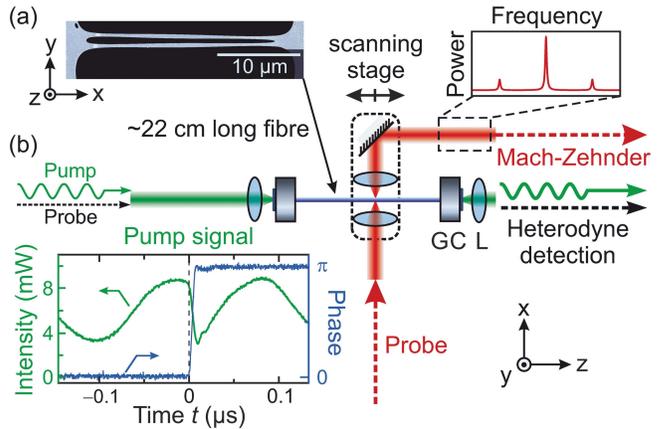

FIG. 5. (a) SEM of the dual-nanoweb fiber structure. (b) Set-up. A ~22-cm-long fiber is mounted between gas cells (GC) and evacuated. Two-frequency pump light at ~1550 nm (green line) and monochromatic probe light at 1543 nm (black dotted line), both propagating along the fiber axis, are coupled in and out with lenses (L). Transverse probe light at 633 nm (red), launched through the cladding into the nanowebs, is used to monitor the nanoweb deflection. Left inset: temporal traces of intensity (green) and phase (blue) of the two-frequency drive signal whose phase is flipped at $t = 0$. Right inset: optical spectrum of transversely transmitted probe light, phase-modulation side-bands generated symmetrically around the carrier by flexural vibrations.

The refractive index modulation induced by flexural vibrations will also generate side-bands at frequencies symmetrically above and below the frequency of a probe laser (right-hand inset in Fig. 5(b)). Their powers scale with phonon population and interaction length. The signals are strongest for probe light at 1543 nm propagating in axial direction, when the phase-modulation accumulates over the whole fiber length. The vibrations can alternatively be imaged with high axial resolution, at the cost of much weaker signals, by side-probing. TE-polarized light at 633 nm was focused onto the nanowebs through the cladding to propagate along them in the $x$-direction. A Mach-Zehnder interferometer converted the phase-modulation of the transmitted light into intensity-modulation at the flexural frequency. This was detected with a high-gain PD and an RF-SA. The minimum resolvable peak deflection of the nanowebs is inversely proportional to the electronically set resolution bandwidth (RBW), over which the power spectrum of electronic noise is integrated. Note that due to the gas cells a 6 cm-long fiber section at each end remains unaccessible for side-probing.

## IV. EXPERIMENTAL RESULTS

Initially we investigated the temporal evolution of the SRLS sidebands for a single step of $\pi$ in the pump-Stokes phase, at 6 mW combined power (below the self-oscillation threshold) and ~1.05 mW total transmitted power. The temporal response of the pump and Stokes signals, monitored using heterodyne detection, is shown in Fig. 6(a). In the steady state (for $t < 0$), the transmitted pump power is depleted by ~150 μW, amplifying the Stokes seed from ~105 to 190 μW. The presence of a population of coherent phonons that can frequency-upshift pump photons is reflected in the emergence of a weak anti-Stokes signal (Fig. 6(b)) following the temporal dynamics of the phonon population.

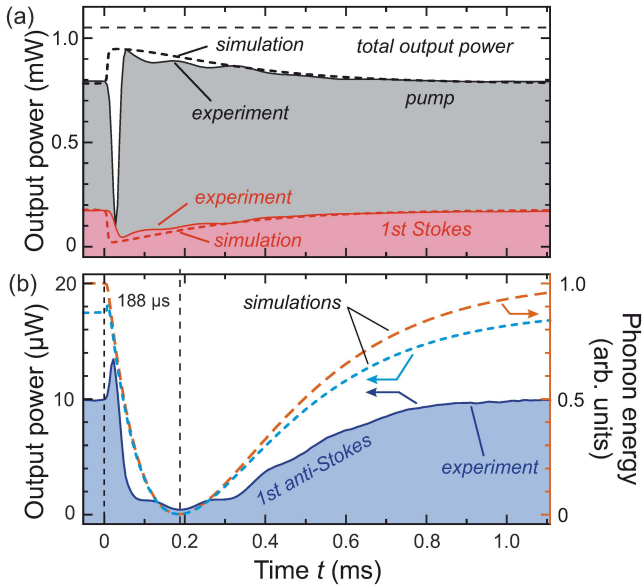

FIG. 6. (a) Heterodyne temporal traces of output pump (black solid line) and Stokes powers (red solid line) for ~6 mW two-frequency input power. Inversion of the relative phase at $t = 0$ causes power-flow from Stokes to pump. (b) Temporal evolution of the SRLS anti-Stokes power (blue solid line). Its signal is smallest at $t \sim 188$ μs. The orange dashed line depicts the simulated evolution of the normalized phonon energy. The other dashed lines in (a) and (b) are numerical simulations of the output power.

When the relative phase is flipped by $\pi$ (at $t = 0$), power is transferred from Stokes to pump (see Fig. 6(a)). This effect is strongest at $t = 0$, decaying as the steady state is again approached. The anti-Stokes power first drops to a minimum at $t = t_{min} = 188$ μs, when destructive interference between phonons generated before and after $t = 0$ is strongest. From Eq.(9) this suggests an average SRLS lifetime $\tau = 188/(2 \ln 2) = 136$ μs. Over the next ~800 μs the anti-Stokes power recovers while the vibrations reach steady state in opposite phase. We attribute the ~40 μs-wide dip in the pump signal at $t = 0$ to the non-instantaneous phase-shift of the pump signal at the PM. This broadens its linewidth beyond the RBW of heterodyne detection, which measures only a fraction of its power. We attribute the weak simultaneous peak in the anti-Stokes signal to interference with the high-frequency tail of the broadened pump line.

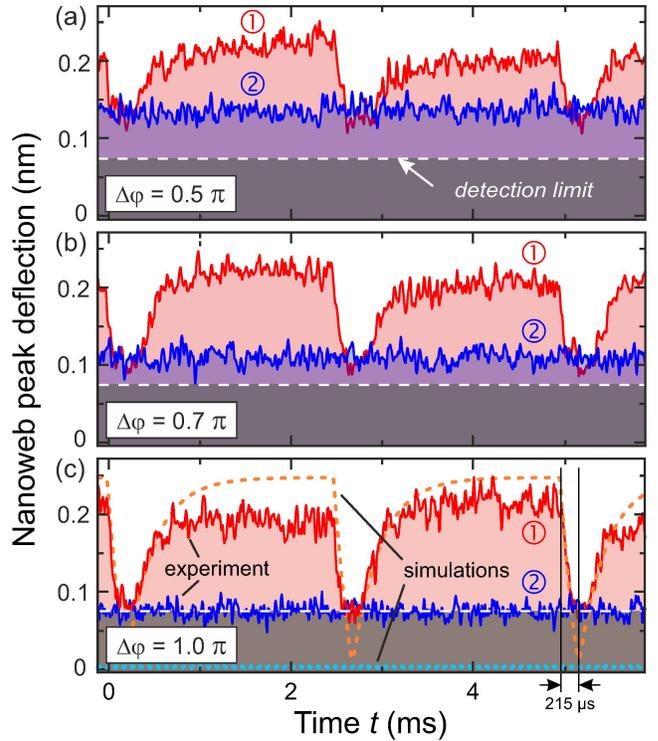

FIG. 7. Temporal evolution of the nanoweb peak deflection, probed from the side of the fiber (solid lines) and numerically simulated (dotted lines). The phase of the SRLS pressure is sequentially switched by (a) $\Delta\varphi = 0.5\pi$, (b) $0.7\pi$ and (c) $\pi$ every 2.5 ms (in red, labelled "①") and 12.5 μs (in blue, labelled "②"). The white dashed line marks the detection limit of side-probing.

Numerical simulations for SRLS in a uniform dual-nanoweb structure (Section II B) support the measured temporal dynamics of the side-bands; although slightly over-estimating the anti-Stokes power, they agree well with the experiment for an effectively 0.8 mm-long fiber (shorter than its geometrical length) over which the SRLS gain peaks. The cumulative energy of the SRLS phonon population along the whole fiber, inaccessible by experimental means, can be also extracted from these simulations. Its temporal trace (orange curve in Fig. 6(b)), normalized to unity at $t = 0$, follows the same trend as the anti-Stokes signal. We attribute the weak ripples in the measured pump and Stokes traces, and the several local minima in the anti-Stokes signal, to transient detuning between the driving beat-note and the vibrations, caused by structural non-uniformities along the fiber. This also means that clean SRLS signals cannot be obtained for switching times less than ~1 ms.

No such restriction exists for side-probing since the spot-size of the focused probe light (~5 µm) is much smaller than the length-scale of the non-uniformities (~1 mm [25]). For a good trade-off between high temporal resolution and detection limit (here ~75 pm), we probed the nanoweb vibrations at 43 mm away from the fiber input, where the steady-state peak deflection was ~220 pm. Figure 7 plots the response of the nanoweb vibrations to periodic phase-steps in the SRLS pressure. When the period is longer than the phonon lifetime (2.5 ms), the vibrations drop to a minimum $t_{min} \sim 200$ µs after each phase-switch, recovering as the phase catches up with SRLS pressure. The minimum deflection falls with increasing depth of the steps, from ~135 pm at $\Delta\varphi = 0.5\pi$ to ~109 pm at $0.7\pi$. The curve measured for $\pi$ is cut off at the detection limit; numerical simulations with $P_{IN} = 3$ mW and $\tau = 160$ µs, agreeing otherwise well with experiments and Eq.(9), predict a minimum deflection of ~2 pm after $2\tau \ln2 = 215$ µs. Switching the SRLS pressure at a period shorter than the phonon lifetime (12.5 µs) causes the nanoweb peak deflection to reset within time $t_{min}$ to a value that matches the minimum observed for the same $\Delta\varphi$ at the longer switching period.

We now analyze the full dynamic range over which the vibrational amplitude can be coherently controlled in the experiment. To this end, we take advantage of the much larger phase-modulation in axial transmission compared to side-probing. Figure 8(a) shows the heterodyne spectrum of the transmitted probe light at 1543 nm. At constant SRLS pressure (no phase steps), the sideband powers are 0.2 µW for Stokes and 2 µW for anti-Stokes when launching 2 mW. We attribute this ~10 dB imbalance to simultaneous generation of SIMS and SRLS phonons when their frequencies coincide. This causes SIMS to transfer probe power to a Stokes signal in the high-loss double-lobed optical mode. Coherent control by flipping the phase of the SRLS pressure every 12.5 µs generates "notches" in both phase-modulation sidebands exactly at the SRLS frequency. These are caused by the interplay of two effects. Firstly, the increase in pump bandwidth during each phase-step increases the beat-note bandwidth, simultaneously driving additional inhomogeneously broadened flexural resonances. Secondly, after each phase-step the pump line narrows again, so that only phonons at $\Omega_R$ are coherently suppressed, resulting in the appearance of a notch. Figure 8(b) plots the depth of this notch versus $\Delta\varphi$, extracted from the anti-Stokes side-band. It is deepest (~24.8 dB) for $\Delta\varphi = \pi$ where the heterodyne signal reaches the detection limit, indicating almost complete suppression of the vibrations along the whole fiber. These results agree well with side-probing and numerical simulations.

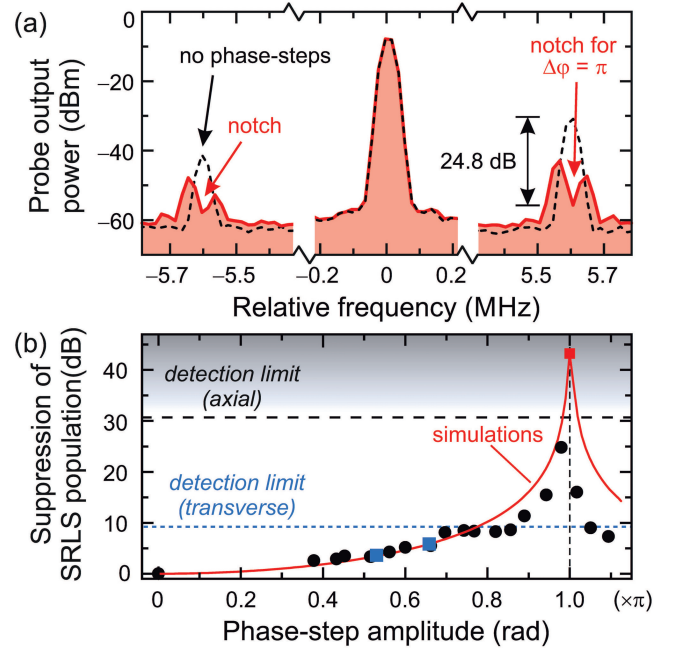

FIG. 8. (a) Heterodyne spectra of axial probe light, phase-modulated by flexural vibrations. Constant SRLS pressure yields largest side-band powers (black dashed line). Notches at the SRLS frequency appear when the SRLS pressure flips every 12.5 µs (red line). (b) Suppression of SRLS phonon population versus phase-step amplitude (black circles: notch depth in anti-Stokes side-band, blue squares: transverse probing, dashed/dotted lines: detection limits, red solid line: numerical simulations).

## V. CONCLUSIONS

Coherent control offers a simple means of adjusting and maintaining the vibrational amplitude of Raman-like mechanical oscillators. We predict and demonstrate experimentally that SRLS uniquely enables simultaneous control and suppression of flexural vibrations over the whole length of a dual-nanoweb fiber. A fast sequence of adjustable steps in the relative phase of a two-frequency signal allows to tune the peak deflection without changing the drive power. Our measurements demonstrate almost 25 dB dynamic range over which the vibrational amplitude can be adjusted on demand. Interestingly, this approach is robust against structural non-uniformities and can be applied to a wide variety of optomechanical systems supporting SRLS [23], including silicon nanowire waveguides [20,21,27], even in presence of thermal effects and nonlinear absorption causing power-dependent optical loss.

# APPENDIX: ROUTE TOWARDS COHERENT CONTROL OF FLEXURAL PHONONS ON FASTER TIMESCALE

The finite time $t_{min}$ for the vibrations to reach their minimum can be viewed as a consequence of their inertia. Nevertheless, in practice this time interval can be manipulated and, in particular, drastically reduced by abruptly increasing the launched power of the two-frequency light by a factor $\xi$ ("beat-note asymmetry"), synchronized to each phase-flip of the beat-note. Under these circumstances, the SRLS pressure is expressed as

$$\Phi(z,t) = \Phi_0(z) - [\Phi_0(z) + \Phi_1(z)]\Theta(t), \quad (A1)$$

where $\Phi_0$ and $-\Phi_1$ are its values before and after $t = 0$, and $\xi = \Phi_1/\Phi_0$. Inserting Eq.(A1) into Eq.(7), the slowly-varying amplitude of SRLS vibrations becomes

$$R(z,t,\xi) = 2i\kappa_R \Phi_0(z)\tau\left[(1+\xi)e^{-t/(2\tau)} - \xi\right]. \quad (A2)$$

Hence, the vibrations reach their minimum after time $t_{min}(\xi) = 2\tau \ln(1 + 1/\xi)$. This time interval equals to $2\tau \ln 2$ for $\xi = 1$ and approaches zero in the limit $\xi \to \infty$. To demonstrate this proposed scheme experimentally, we employ an electro-optic intensity modulator (EOM) at the output of the laser system to control the power of the two-frequency light. The EOM is driven by rectangular voltage pulses with a duration $\tau_{pulse} \approx 150$ μs and a period of 5 ms, their rising edges synchronized to the phase-steps in the SRLS pressure.

Figure 9 plots temporal traces of the nanoweb peak deflection, transversely probed at (a) 43 and (b) 47 mm away from the fiber input. When $\xi$ is consecutively increased, the time interval $t_{min}$ is found to shrink simultaneously at both positions. To avoid uncertainties in the quantitative evaluation due to the broad and noisy minima, we define an effective suppression time $t_{sup}$ for the vibrations to reach a normalized peak deflection $r_{ref}$, whose RF power in transverse probing is 3 dB above noise. This level, indicated by a dashed line, is 44 pm in Fig. 9(a) and 50 pm in Fig. 9(b). The data-points are averaged for each value of ξ over ten consecutive traces (one of which is shown in Fig. 9(a),(b)) and plotted in Fig. 9(c),(d). They are found to agree well with the theoretical curve given by

$$t_{\sup} = 2\tau \ln\left(1 + \frac{1 - r_{\text{ref}}}{\xi + r_{\text{ref}}}\right), \tag{A3}$$

which are shown as black solid lines. To keep the change of the average optical power as small as possible, we limited the power increase to a duration $\tau_{\text{pulse}}$ much shorter than the phase-step period.

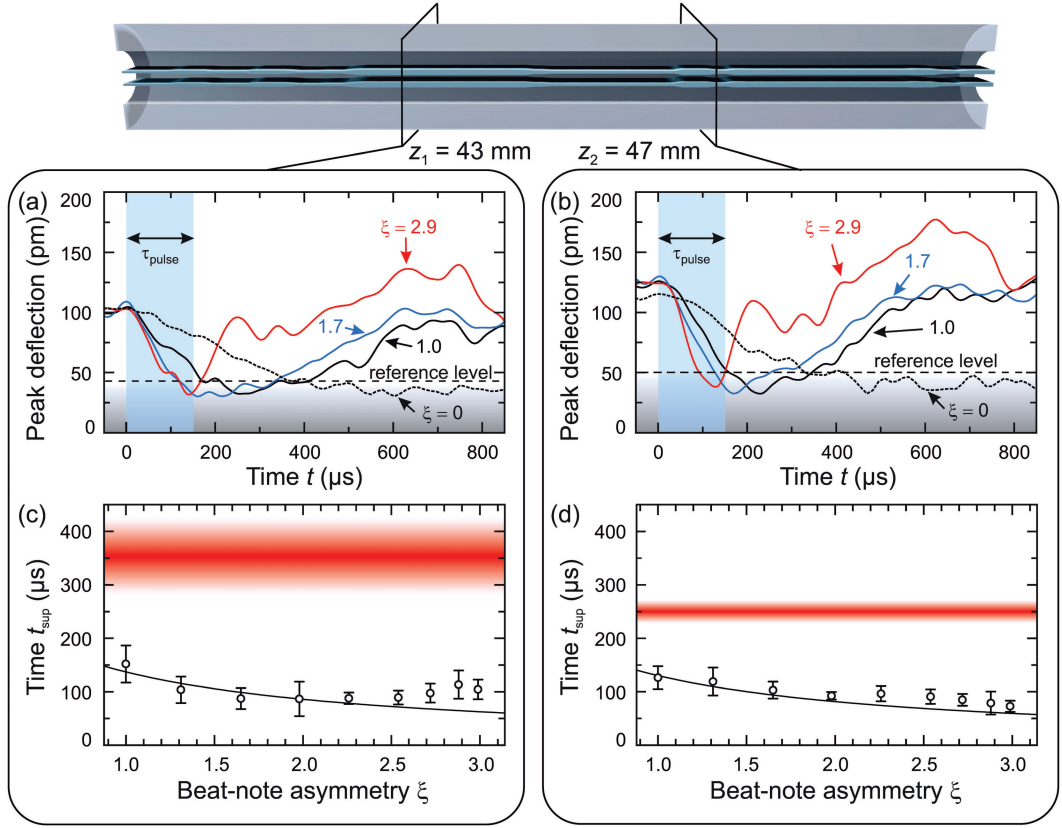

FIG. 9. Reduction of $t_{sup}$ when the phase-steps in the SRLS pressure are synchronized with a 150 μs-long increase of the two-frequency power. Temporal traces of the nanoweb peak deflection were probed at (a) 43 and (b) 47 mm away from the input fiber end (dotted lines: ring-down, dashed lines: reference levels). The suppression time $t_{sup}$ is plotted in (c), (d) versus beat-note asymmetry. Open circles are (averaged) data points, error bars their standard deviation, and Eq.(A3) is shown as a solid-black line. Red lines are averages of six ring-downs, their standard deviation indicated by shading.